\documentclass[a4paper,fleqn,usenatbib,useAMS]{mnras}

\usepackage{graphicx}	
\usepackage{amsmath}	
\usepackage{amssymb}	
\usepackage{multicol}        
\usepackage{bm}		

\newcommand{\Omegadot}{\dot{\Omega}}
\newcommand{\tmag}{t_{\rm mag}}
\newcommand{\tdiff}{t_{\rm diff}}
\newcommand{\tpeak}{t_{\rm peak}}
\newcommand{\Erot}{E_{\rm rot}}
\newcommand{\Edotrot}{\dot{E}_{\rm rot}}
\newcommand{\Edotmag}{\dot{E}_{\rm mag}}
\newcommand{\Edotgw}{\dot{E}_{\rm gw}}
\newcommand{\Edotgwr}{\dot{E}_{\rm gw,r}}
\newcommand{\Lmag}{L_{\rm mag}}

\newcommand{\Lrad}{L_{\rm rad}}
\newcommand{\Lpeak}{L_{\rm peak}}
\newcommand{\Omegapeak}{\Omega_{\rm peak}}
\newcommand{\gammae}{\gamma_{\rm e}}
\newcommand{\gammar}{\gamma_{\rm r}}
\newcommand{\tgw}{t_{\rm gw}}
\newcommand{\tgwr}{t_{\rm gw,r}}
\newcommand{\Esn}{E_{\rm SN}}
\newcommand{\Mej}{M_{\rm ej}}

\usepackage[T1]{fontenc}
\usepackage{ae,aecompl}

\usepackage{txfonts}

\title[GWs within magnetar model of SLSNe and GRBs]{Gravitational waves within
the magnetar model of superluminous supernovae and gamma-ray bursts}

\author[W. C. G. Ho]{Wynn C. G. Ho\thanks{Email: \href{mailto:wynnho@slac.stanford.edu}{wynnho@slac.stanford.edu}}
\\
Mathematical Sciences, Physics and Astronomy and STAG Research Centre, University of Southampton, Southampton, SO17 1BJ, UK \\
}

\date{Accepted 2016 August 9. Received 2016 June 17; in original form 2016 May 23}

\pubyear{2016}

\begin{document}
\label{firstpage}
\pagerange{\pageref{firstpage}--\pageref{lastpage}}
\maketitle

\begin{abstract}
The light curve of many supernovae (SNe) and gamma-ray bursts (GRBs) can be
explained by a sustained injection of extra energy from its possible central
engine, a rapidly rotating strongly magnetic neutron star (i.e. magnetar).
The magnetic dipole radiation power that the magnetar supplies comes at the
expense of the star's rotational energy.
However, radiation by gravitational waves (GWs) can be more efficient than
magnetic dipole radiation because of its stronger dependence on neutron star
spin rate $\Omega$, i.e. $\Omega^6$ (for a static `mountain') or $\Omega^8$
(for a r-mode fluid oscillation) versus $\Omega^4$ for magnetic dipole
radiation.
Here, we use the magnetic field $B$ and initial spin period $P_0$ inferred
from SN and GRB observations to obtain simple constraints
on the dimensionless amplitude of the mountain of $\varepsilon<0.01$ and
r-mode oscillation of $\alpha<1$, the former being similar to that obtained
by recent works.
We then include GW emission within the magnetar model.
We show that when $\varepsilon>10^{-4}(B/10^{14}\mbox{ G})(P_0/1\mbox{ ms})$ or
$\alpha>0.01(B/10^{14}\mbox{ G})(P_0/1\mbox{ ms})^2$,
light curves are strongly affected, with significant decrease in peak
luminosity and increase in time to peak luminosity.
Thus the GW effects studied here are more pronounced for low $B$ and short
$P_0$ but are unlikely to be important in modelling SN and GRB light curves
since the amplitudes needed for noticeable changes are quite large.
\end{abstract}


\section{Introduction} \label{sec:intro}

A subset of supernovae (SNe) and gamma-ray bursts (GRBs) requires an additional
source of energy input over an extended period of time
in order to explain the observed evolution of their brightness or light curve.
The magnetar model can provide such an energy and timescale
(see, e.g. \citealt{usov92,dailu98,zhangmeszaros01} for GRBs and
\citealt{maedaetal07,kasenbildsten10,woosley10} for SNe),
and this model has been used quite successfully
(see, e.g. \citealt{rowlinsonetal13,luetal15} for GRBs and
\citealt{chatzopoulosetal13,inserraetal13,nicholletal14} for SNe).
The magnetar model pre-supposes that a rapidly rotating magnetar
(i.e. neutron star with spin period $P\sim 1\mbox{ ms}$ and magnetic field
$B\sim 10^{14}\mbox{ G}$; see, e.g.
\citealt{mereghetti08,mereghettietal15,turollaetal15}, for review of magnetars)
forms during the SN.
The newborn magnetar supplies extra energy to power the SN or GRB light curve
via its large rotational energy
\begin{equation}
\Erot = I\Omega^2/2 = 2.0\times 10^{52}\mbox{ erg }
 (P/1\mbox{ ms})^{-2}, \label{eq:erot}
\end{equation}
where $\Omega$ $(=2\pi/P)$ and $I$ $(\approx 10^{45}\mbox{ g cm$^2$ })$
are the magnetar angular spin frequency and moment of inertia, respectively.
This rotational energy is deposited within the SN on the timescale over
which the magnetar loses energy via magnetic dipole radiation, i.e.
\begin{equation}
\tmag = \left|\frac{\Erot}{\Edotmag}\right|_{\Omega_0}
 = \frac{1}{2\beta\Omega_0^2}
 = 2.0\times 10^{5}\mbox{ s }B_{14}^{-2}(P_0/1\mbox{ ms})^{2}, \label{eq:tmag}
\end{equation}
where $B_{14}=B/10^{14}\mbox{ G}$,
$\Omega_0$ and $P_0$ are initial spin frequency and period, respectively,
$\beta\equiv B^2R^6/6c^3I=6.2\times 10^{-14}\mbox{ s }B_{14}^2$, and
we assume a neutron star mass $M=1.4\,M_{\sun}$, radius $R=10\mbox{ km}$,
and $I=10^{45}\mbox{ g cm$^2$ }$ hereafter.
Magnetic dipole radiation energy loss $\Edotmag$ is the power supplied
by the magnetar $\Lmag$ to the SN and is given by
\begin{eqnarray}
\Lmag &\equiv&\Edotmag=-\frac{B^2R^6\Omega^4\sin^2\theta}{6c^3}=-\beta I\Omega^4
 \nonumber\\
 &=& -9.6\times 10^{46}\mbox{ erg s$^{-1}$ } B_{14}^2
 (P/1\mbox{ ms})^{-4}, \label{eq:lmag}
\end{eqnarray}
where $\theta$ is the angle between stellar rotation and magnetic axes
(\citealt{pacini68,gunnostriker69};
see also \citealt{spitkovsky06,contopoulosetal14}).
For simplicity, we assume an orthogonal rotator, i.e. $\sin\theta=1$.
It is also worth noting that while the theoretical minimum spin period is
$\sim 0.3-0.5\mbox{ ms}$ \citep{cooketal94,korandaetal97,haenseletal99},
the lowest observed radio
pulsar spin period is 1.4~ms \citep{hesselsetal06}.

The millisecond magnetar model for SNe and GRBs is useful because it provides
an easy explanation for the requisite extra energy and timescale.
These last two map directly to the initial magnetar magnetic field and
spin period (see equations~\ref{eq:erot} and \ref{eq:tmag}),
with somewhat weaker dependence on other parameters (see below).
However, the effect of emission of gravitational waves (GWs) merits
consideration because GW energy loss $\Edotgw$ scales with spin
frequency at a higher power than the scaling of magnetic dipole radiation
(see equation~\ref{eq:erot}), and thus the assumption and need for rapid
rotation can lead to $\Edotgw\gtrsim\Lmag$.
GWs from a neutron star can be produced in two ways, by a static
quadrupolar deformation (`mountain') or by a fluid oscillation
(see, e.g. \citealt{lasky15}).
For a mountain with size $\varepsilon$, the GW energy loss rate and
timescale are (see, e.g. \citealt{shapiroteukolsky83})
\begin{eqnarray}
\Edotgw &=& -\frac{32}{5}\frac{GI^2}{c^5}\varepsilon^2\Omega^6
 = -\gammae I\Omega^6 \nonumber\\
 &=& 1.1\times 10^{46}\mbox{ erg s$^{-1}$ }\varepsilon_{-4}^2
 (P/1\mbox{ ms})^{-6} \\
\tgw &=& \left|\frac{\Erot}{\Edotgw}\right|_{\Omega_0}\!\!
 = \frac{1}{2\gammae\Omega_0^4} = 1.8\times 10^{6}\mbox{ s }
 \varepsilon_{-4}^{-2}(P_0/1\mbox{ ms})^{4}, \label{eq:tgw}
\end{eqnarray}
respectively, where $\gammae\equiv 32GI\varepsilon^2/5c^5
=1.8\times 10^{-14}\mbox{ s$^{3}$}\varepsilon^2$
and $\varepsilon_{-4}=\varepsilon/10^{-4}$.
Alternatively, if the neutron star has a r-mode oscillation with amplitude
$\alpha$ \citep{anderssonkokkotas01}, then the
GW energy loss rate and timescale are (see, e.g. \citealt{owenetal98})
\begin{eqnarray}
\Edotgwr\!\!\!\! &\approx& -\frac{96\pi}{15^2}\left(\frac{4}{3}\right)^6
 \frac{GMR^4\tilde{J}^2I}{c^7\tilde{I}}\alpha^2\Omega^8
 = -\gammar I\Omega^8 \nonumber\\
 &=& 1.6\times 10^{46}\mbox{ erg s$^{-1}$ }\alpha_{-2}^2 (P/1\mbox{ ms})^{-8} \\
\tgwr\!\!\!\! &=& \left|\frac{\Erot}{\Edotgwr}\right|_{\Omega_0}\!\!
 = \frac{1}{2\gammar\Omega_0^6} = 1.2\times 10^{6}\mbox{ s }\alpha_{-2}^{-2}
 (P_0/1\mbox{ ms})^{6}, \label{eq:tgwr}
\end{eqnarray}
respectively, where
$\gammar\equiv (96\pi/15^2)(4/3)^6(GMR^4\tilde{J}^2/c^7\tilde{I})\alpha^2
=6.6\times 10^{-26}\mbox{ s$^{5}$}\alpha^2$ and
$\alpha_{-2}=\alpha/10^{-2}$.
Note that $\tilde{J}$ $(=0.01635)$ and $\tilde{I}$ $(=I/MR^2=0.261)$ are derived
using a $\Gamma=2$ polytrope with $R=12.53\mbox{ km}$ \citep{owenetal98},
whereas we take $R=10\mbox{ km}$,
but this difference will not change our results qualitatively.

GW emission by the magnetar has a two-fold effect on its ability
to provide extra power to a SN or GRB: (1) By causing the neutron star spin
rate $\Omega$ to decrease faster than by magnetic dipole radiation
(when $\tgw$ or $\tgwr\lesssim\tmag$; see below), the timescale over which
$\Erot(\Omega)$ is supplied to the SN/GRB is shorter; (2) Since the energy
emitted by GWs is not imparted to the SN/GRB, the time evolution of the
energy supplied to the SN/GRB changes, and thus the predicted light curve
changes.
In Section~\ref{sec:magnetar}, we use (1) and inferred values of initial
spin period $P_0$ and magnetic field $B$ to obtain the simplest constraint
on GW ellipticity $\varepsilon$ and amplitude $\alpha$.
In Section~\ref{sec:magnetargw}, we account for GWs within the magnetar
model of \cite{kasenbildsten10} in order to evaluate the effect on light
curves and compare to the magnetar model without GWs.
We summarize and briefly discuss our results in Section~\ref{sec:discuss}.
We note that several recent works performed similar analysis to that done here.
\citet{moriyatauris16} compare timescales $\tmag$ and $\tgw$ in order to
constrain $\varepsilon$ for superluminous SNe,
while \citet{laskyglampedakis16} do the same for short GRBs;
these works assume that the spin period evolution is determined by
{\it either} magnetic dipole radiation or GW emission due to an ellipticity,
as is done in Section~\ref{sec:magnetar},
whereas their combined effects are considered in Section~\ref{sec:magnetargw}.
\citet{kashiyamaetal16} use a somewhat more detailed model compared to that
presented in Section~\ref{sec:magnetargw} in order to evaluate effects of
GWs (with $\varepsilon$ due to extreme magnetic fields)
on light curves of superluminous SNe
(see also \citealt{zhangmeszaros01} for GRBs).

\section{Magnetar model without GWs: GW constraints}
\label{sec:magnetar}

\begin{figure}
 \includegraphics[width=\columnwidth]{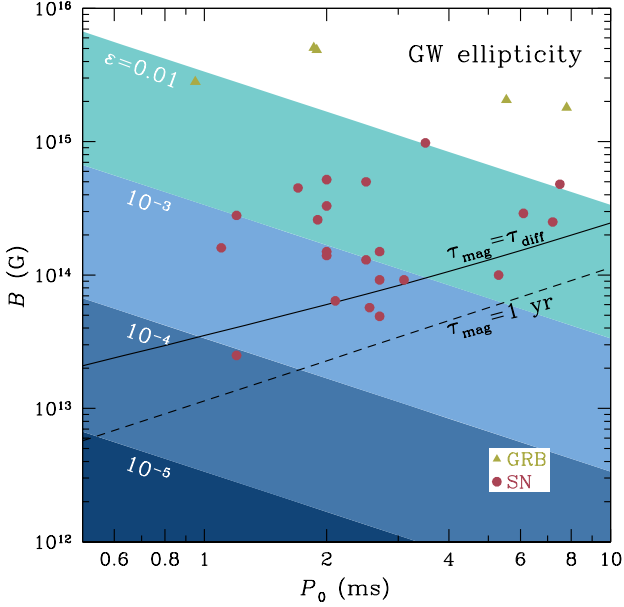}
 \caption{
Constraint on GW ellipticity $\varepsilon$ based on equation~(\ref{eq:epsilon})
as a function of neutron star initial spin period $P_0$ and magnetic field $B$.
Solid line denotes when the magnetic dipole spin-down timescale $\tmag$
(see equation~\ref{eq:tmag}) is equal to the radiative diffusion timescale
$\tdiff$ (see equation~\ref{eq:tdiff}),
and the dashed line denotes when $\tmag$ is equal to 1~yr.
Circles are $P_0$ and $B$ for SNe
(see \citealt{moriyatauris16}, and references therein)
and triangles are $P_0$ and $B$ for GRBs
(see \citealt{laskyetal14}, and references therein).
}
 \label{fig:ellip}
\end{figure}

\begin{figure}
 \includegraphics[width=\columnwidth]{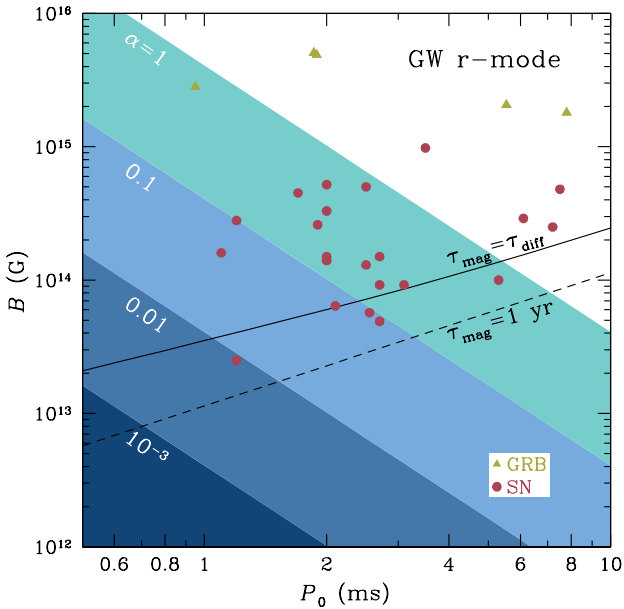}
 \caption{
Constraint on GW r-mode amplitude $\alpha$ based on equation~(\ref{eq:alpha})
as a function of neutron star initial spin period $P_0$ and magnetic field $B$.
Solid line denotes when the magnetic dipole spin-down timescale $\tmag$
(see equation~\ref{eq:tmag}) is equal to the radiative diffusion timescale
$\tdiff$ (see equation~\ref{eq:tdiff}),
and the dashed line denotes when $\tmag$ is equal to 1~yr.
Circles are $P_0$ and $B$ for SNe
(see \citealt{moriyatauris16}, and references therein)
and triangles are $P_0$ and $B$ for GRBs
(see \citealt{laskyetal14}, and references therein).
}
 \label{fig:rmode}
\end{figure}

In this section, we consider the simplest GW effect on the magnetar model.
We assume that some SN and GRB light curves are well-explained by the
magnetar model {\it without} GWs.
As a result of fits to these light curves, the magnetar initial spin period
$P_0$ and magnetic field $B$ are extracted (see, e.g.
\citealt{maedaetal07,trojaetal07,chatzopoulosetal13,inserraetal13,rowlinsonetal13,nicholletal14,luetal15}).
For the neutron star spin rate to decrease by only magnetic dipole radiation,
the timescale for this energy loss must be shorter than that due to GWs, i.e.
$\tmag<\tgw$ or
\begin{equation}
\tmag/\tgw = \gammae\Omega_0^2/\beta = 0.11\,B_{14}^{-2}\varepsilon_{-4}^2
 (P_0/1\mbox{ ms})^{-2}<1 \label{eq:tmagtgw}
\end{equation}
and $\tmag<\tgwr$ or
\begin{equation}
\tmag/\tgwr = \gammar\Omega_0^4/\beta
 =0.17\,B_{14}^{-2}\alpha_{-2}^2 (P_0/1\mbox{ ms})^{-4}<1. \label{eq:tmagtgwr}
\end{equation}
Thus the constraint on GW ellipticity $\varepsilon$ and r-mode amplitude
$\alpha$ are
\begin{eqnarray}
\varepsilon &<& 3.0\times 10^{-4}\,B_{14}(P_0/1\mbox{ ms}) \label{eq:epsilon} \\
 \alpha &<& 0.025\,B_{14}(P_0/1\mbox{ ms})^{2} \label{eq:alpha},
\end{eqnarray}
respectively,
because otherwise the neutron star spin rate would decrease more quickly
than it would in the magnetar model without GWs.
Equation~(\ref{eq:epsilon}) is identical to that found in
\citet{moriyatauris16}, and thus our constraints on $\varepsilon$ are the
same as theirs.
\citet{laskyglampedakis16} derive a nearly identical constraint equation
for $\varepsilon$ but include an efficiency factor which leads to
constraints about ten times smaller.
Our constraints on $\varepsilon$ and $\alpha$ are shown in
Figs~\ref{fig:ellip} and \ref{fig:rmode},
alongside inferred $P_0$ and $B$ compiled by \citet{moriyatauris16} for SNe
and \citet{laskyetal14} for GRBs;
note that these inferred spin period and magnetic field values are derived
from fits of observed light curves using a model that assumes no contribution
by GW energy loss.
We see that ellipticity is constrained to be $\varepsilon<10^{-3}$ and
r-mode amplitude is constrained to be $\alpha\lesssim 0.1$ for many SNe.
Because of its low inferred magnetic field (thus weaker magnetic dipole
radiation loss) and fast inferred rotation rate, ASASSN-15lh has the
strongest constraints of $\varepsilon<10^{-4}$ and $\alpha<10^{-2}$.
While sources with lower $B$ and/or $P_0$ provide stronger GW constraints,
the magnetar contribution to the SN/GRB light curve
(i.e. $\Lmag$; see equation~\ref{eq:lmag}) decreases with lower magnetic
field, such that shorter spin period is preferred.

\section{Magnetar model with gravitational waves}
\label{sec:magnetargw}

Here we build on the magnetar model of \citet{kasenbildsten10} (see also
\citealt{ostrikergunn71,arnett79,arnett80,metzgeretal15}) by accounting for the effect of GW
emission by the magnetar (see also \citealt{muraseetal15,kashiyamaetal16}).
If the neutron star is a strong enough emitter of GWs, then this energy
loss will cause the star's spin frequency to decrease at a faster rate
than by pure magnetic dipole radiation.  As a consequence, the amount of
rotational energy that can be supplied to the SN or GRB is reduced, and
the shape of the light curve, e.g. peak luminosity and decay rate, will
be altered from that predicted by a model which only considers magnetic
dipole radiation.

At early times ($t\lesssim\tdiff$, where $\tdiff$ is the photon diffusion
timescale; see below) when the SN or GRB is opaque to photons, the radiated
luminosity $\Lrad$ (or photon energy loss) is that given by radiative
diffusion, i.e.
\begin{eqnarray}
\Lrad &=&
 4\pi r^2\left[\frac{c}{3\kappa\rho}\frac{\upartial(E/V)}{\upartial r}\right]
 \approx \frac{4\pi cv}{3\kappa\Mej}tE = \frac{tE}{\tdiff^2} \nonumber\\
 &=& 10^{46}\mbox{ erg s$^{-1}$}\left(\frac{\tdiff}{10^6\mbox{ s}}\right)^{-2}
 \left(\frac{E}{10^{52}\mbox{ erg}}\right)\left(\frac{t}{10^6\mbox{ s}}\right),
 \label{eq:Lrad}
\end{eqnarray}
where $E$ is thermal energy, $V$ is volume, $\kappa$ is opacity,
$\rho$ is density, $r=vt$, ejecta mass is $\Mej=\rho V$,
and expansion velocity is $v\approx\{[\Erot(\Omega)+\Esn]/\Mej\}^{1/2}$,
where $\Esn$ is the initial explosion energy of the SN or GRB.
The photon or radiative diffusion timescale is (see, e.g. \citealt{arnett79})
\begin{eqnarray}
\tdiff &=& \left(\frac{3\kappa\Mej}{4\pi cv}\right)^{1/2}
= \left[\frac{3\kappa\Mej^{3/2}}{4\pi c(\Erot+\Esn)^{1/2}}\right]^{1/2}
 \nonumber\\
 &=& 2.1\times 10^6\mbox{ s}
 \left(\frac{\kappa}{0.2\mbox{ cm$^2$ g$^{-1}$}}\right)^{1/2}
 \left(\frac{\Mej}{1\,M_\odot}\right)^{3/4}
 \left(\frac{10^{51}\mbox{ erg}}{\Erot+\Esn}\right)^{1/4}. \label{eq:tdiff}
\end{eqnarray}

To determine the light curve or evolution of $\Lrad$, we solve the energy
equation (see, e.g. \citealt{arnett79,arnett80})
\begin{equation}
\frac{\upartial E}{\upartial t} = -p\frac{\upartial V}{\upartial t}+\Lmag-\Lrad.
 \label{eq:energy}
\end{equation}
The first term on the right-hand side is energy lost to expansion due to
pressure $p$,
the second is energy supplied by the rapidly rotating magnetar,
and the third is energy radiated as photons.
Note that energy radiated as GWs is accounted for in the lower energy
provided by the magnetar.
We could also include a term due to heating by Ni decay
$L_{\rm Ni}\sim 10^{43}\mbox{ erg s$^{-1}$}e^{-t/8.8\mbox{ d}}$
\citep{metzgeretal15,kashiyamaetal16};
however, this would not significantly affect results presented here since we
consider rapid rotation, such that $\Lmag\gg L_{\rm Ni}$.
When pressure is dominated by radiation, such that $p=(1/3)E/V$,
then $p(\upartial V/\upartial t)=E/t$, and equation~(\ref{eq:energy}) becomes
\begin{eqnarray}
\frac{1}{t}\frac{\upartial(tE)}{\upartial t} &=& \Lmag-\Lrad \nonumber\\
\frac{\upartial(tE)}{\upartial t}
 &=& \beta I\Omega^4t-\frac{(tE)}{\tdiff^2(\Omega)}t. \label{eq:edot}
\end{eqnarray}
Evolution of the neutron star rotation rate is obtained from
$\Edotrot=\Edotmag+\Edotgw$, where
$\Edotrot = I\Omega\Omegadot
 = 3.9\times 10^{54}\mbox{ erg s$^{-1}$}(P/1\mbox{ ms})^{-3} \dot{P}$,
where $\Omegadot$ is time derivative of $\Omega$.
For simplicity, we neglect possible accretion on to the newborn neutron star,
the effect of which depends on accretion rate and could spin-up or spin-down
the star (see, e.g. \citealt{piroott11,melatospriymak14}).
For GWs from an ellipticity $\varepsilon$, the evolution equation for spin
frequency is then (see, e.g. \citealt{shapiroteukolsky83})
\begin{equation}
\frac{d\Omega}{dt} = -\beta\Omega^3-\gammae\Omega^5. \label{eq:omegadot}
\end{equation}
This can be solved analytically to yield \citep{ostrikergunn69}
\begin{equation}
t = \int_{\Omega}^{\Omega_0}\!\!\!\!
 \frac{d\Omega}{\Omega^3(\beta+\gammae\Omega^2)}
 = \left[\frac{\gammae}{2\beta^2}
 \ln\left(\frac{\beta+\gammae\Omega^2}{\Omega^2}\right)-\frac{1}{2\beta\Omega^2}
 \right]_{\Omega}^{{\Omega_0}}
\end{equation}
and, after some algebra, 
\begin{equation}
\frac{\Omega^2}{1-\frac{\gammae\Omega^2}{\beta}
 \ln\left(\frac{1+\beta/\gammae\Omega^2}{1+\beta/\gammae\Omega_0^2}\right)}
 = \frac{\Omega_0^2}{1+2\beta\Omega_0^2t}. \label{eq:omegae}
\end{equation}
Analogous to equation~(\ref{eq:omegadot}), the spin evolution equation for
GW from a r-mode is
\begin{equation}
\frac{d\Omega}{dt} = -\beta\Omega^3-\gammar\Omega^7, \label{eq:omegadotr}
\end{equation}
with a solution that is given by
\begin{equation}
\frac{\Omega^2}{1-\left(\frac{\gammar\Omega^4}{\beta}\right)^{1/2}\tan^{-1}
 \left[\frac{(\gammar\Omega_0^4/\beta)^{1/2}(1-\Omega^2/\Omega_0^2)}{1+\gamma\Omega_0^2\Omega^2/\beta}\right]}
 = \frac{\Omega_0^2}{1+2\beta\Omega_0^2t}. \label{eq:omegar}
\end{equation}
For equations~(\ref{eq:omegae}) and (\ref{eq:omegar}),
it is clear that when there is no additional torque (such as that due to
GWs), i.e. $\gammae=0$ and $\gammar=0$, the spin frequency
evolution $\Omega(t)$ reduces to that due to a magnetic dipole
$\Omega(t)=\Omega_0(1+2\beta\Omega_0^2t)^{-1/2}=\Omega_0(1+t/\tmag)^{-1/2}$
and substituting into equation~(\ref{eq:lmag}) yields
$\Lmag(t)=-9.6\times 10^{46}\mbox{ erg s$^{-1}$}B_{14}^2(P_0/1\mbox{ ms})^{-4}
(1+t/\tmag)^{-2}$.

\begin{figure}
 \includegraphics[width=\columnwidth]{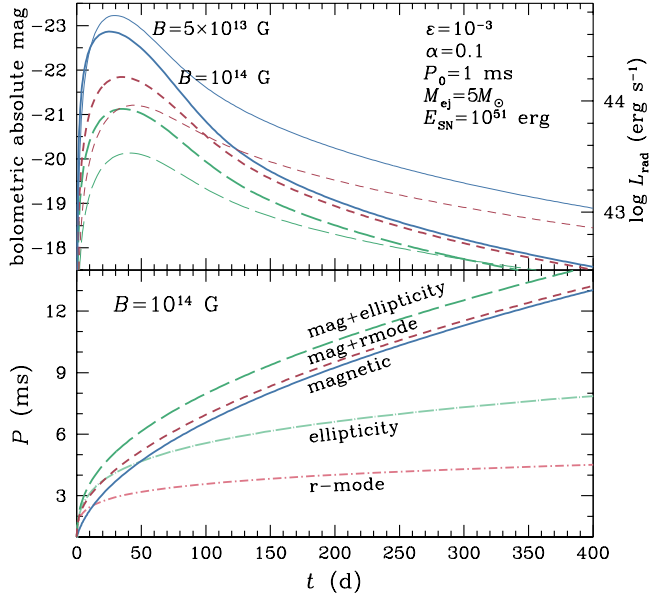}
 \caption{
Light curves of radiated luminosity $\Lrad$ and spin period $P$ as a function
of time.
Solid lines are for a model which only includes magnetic dipole energy loss
$\Lmag$, with parameters $P_0=1\mbox{ ms}$,
$M_{\rm ej}=5M_{\sun}$, $\Esn=10^{51}\mbox{erg}$, and either
$B=10^{14}\mbox{ G}$ (bold lines) or $B=5\times 10^{13}\mbox{ G}$
(light lines).
Long-dashed lines are for a model which includes magnetic dipole and GW energy
loss $\Edotgw$, with ellipticity $\varepsilon=10^{-3}$.
Short-dashed lines are for a model which includes magnetic dipole and GW energy
loss $\Edotgwr$, with r-mode amplitude $\alpha=0.1$.
Dot-dashed lines are for spin period evolution by only GW ellipticity or r-mode.
}
 \label{fig:evol1}
\end{figure}

\begin{figure}
 \includegraphics[width=\columnwidth]{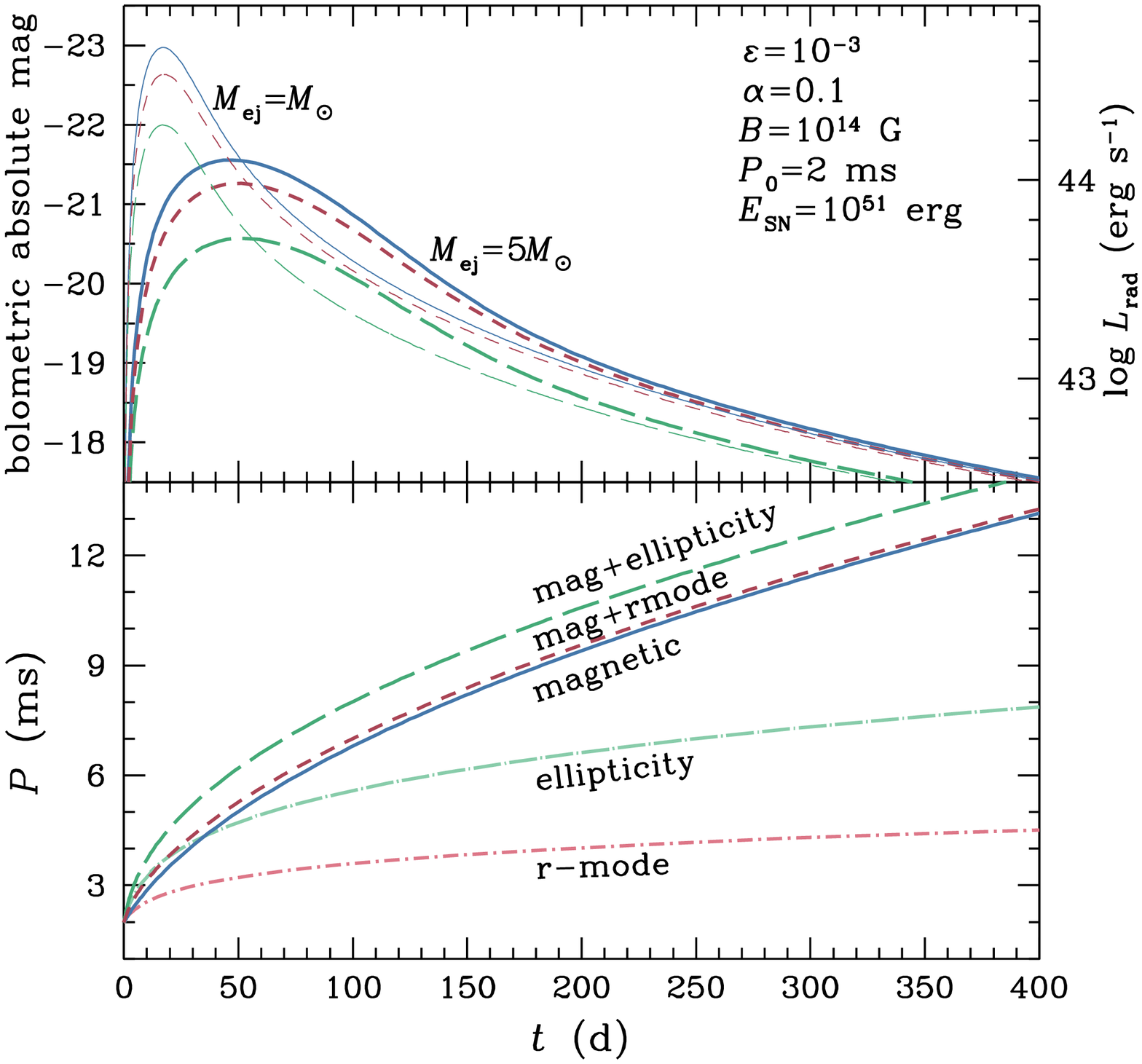}
 \caption{
Light curves of radiated luminosity $\Lrad$ and spin period $P$ as a function
of time.
Solid lines are for a model which only includes magnetic dipole energy loss
$\Lmag$, with parameters $B=10^{14}\mbox{ G}$, $P_0=2\mbox{ ms}$,
$\Esn=10^{51}\mbox{erg}$, and either $M_{\rm ej}=5M_{\sun}$ (bold lines) or
$M_{\rm ej}=M_{\sun}$ (light lines).
Long-dashed lines are for a model which includes magnetic dipole and GW energy
loss $\Edotgw$, with ellipticity $\varepsilon=10^{-3}$.
Short-dashed lines are for a model which includes magnetic dipole and GW energy
loss $\Edotgwr$, with r-mode amplitude $\alpha=0.1$
Dot-dashed lines are for spin period evolution by only GW ellipticity or r-mode.
}
 \label{fig:evol2}
\end{figure}

Examples of solutions of the
coupled\footnote{It is simple to solve either differential equations for
$\Omega$ rather than use the analytic solution of $\Omega(t)$ given by
equations~(\ref{eq:omegae}) or (\ref{eq:omegar}).}
evolution equations~(\ref{eq:edot}) and (\ref{eq:omegadot}) with
$\varepsilon=10^{-3}$ or equations~(\ref{eq:edot}) and (\ref{eq:omegadotr})
with $\alpha=0.1$ are shown in Figs~\ref{fig:evol1} and \ref{fig:evol2}.
Initial spin period is $P_0=1\mbox{ ms}$ for Fig.~\ref{fig:evol1}
and 2~ms for Fig.~\ref{fig:evol2}.
Results for two magnetic fields ($B=5\times 10^{13}$ and $10^{14}\mbox{ G}$)
are shown in Fig.~\ref{fig:evol1}, while two ejecta masses
($\Mej=M_{\sun}$ and $5M_{\sun}$) are shown in Fig.~\ref{fig:evol2}.
Bottom panels also show spin period evolution due to only a
GW ellipticity or r-mode, i.e. $P=P_0(1+2t/\tgw)^{1/4}$ or
$P=P_0(1+3t/\tgwr)^{1/6}$, respectively, compared to
$P=P_0(1+t/\tmag)^{1/2}$ for only magnetic dipole radiation
(see after equation~\ref{eq:omegar}).
GW emission determines the spin evolution at early times
($t\lesssim\mbox{ several days}$), while magnetic dipole radiation dominates
at late times
so that $\Lmag\propto(1+t/\tmag)^{-2}$ (see after equation~\ref{eq:omegar}).

\begin{figure}
 \includegraphics[width=\columnwidth]{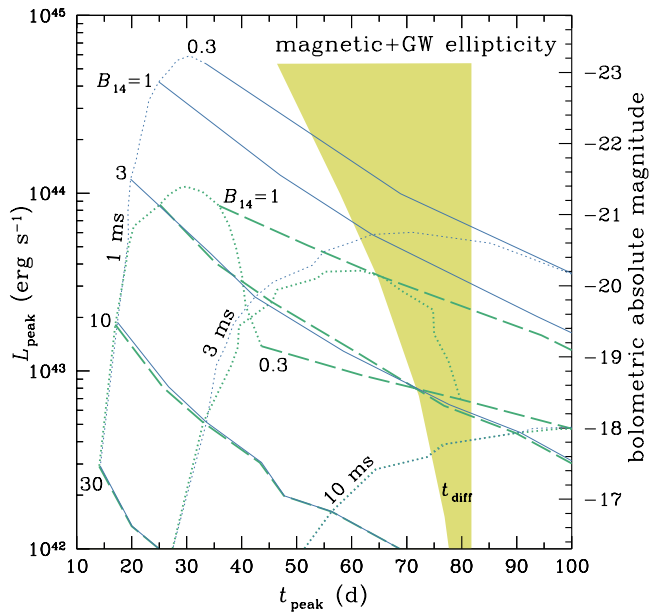}
 \caption{
Parameter space of peak radiated photon luminosity $\Lpeak$ and time to peak
luminosity $\tpeak$.
Light solid lines are for a model which only includes magnetic dipole energy
loss, where magnetic field $B=10^{14}B_{14}$ (labelled) is held
constant and initial spin period $P_0$ is varied,
while light dotted lines are for constant $P_0$ (labelled) and varying $B$.
Heavy dashed and dotted lines are the same but for a model which includes
magnetic dipole and GW energy loss, with ellipticity $\varepsilon=10^{-3}$.
Shaded region shows range of diffusion time which varies because
$\tdiff=\tdiff[\Erot(\Omegapeak)]$.
All models shown assume $\Esn=10^{51}\mbox{erg}$ and $M_{\rm ej}=5M_{\sun}$.
}
 \label{fig:peake}
\end{figure}

\begin{figure}
 \includegraphics[width=\columnwidth]{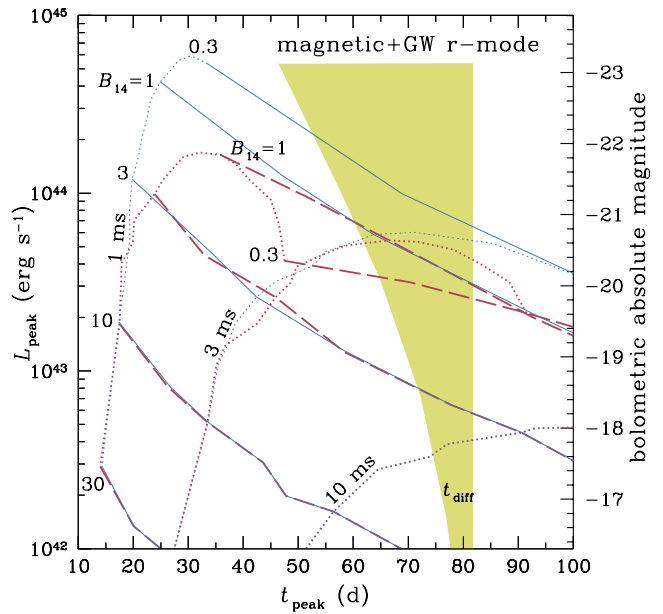}
 \caption{
Parameter space of peak radiated photon luminosity $\Lpeak$ and time to peak
luminosity $\tpeak$.
Light solid lines are for a model which only includes magnetic dipole energy
loss, where magnetic field $B=10^{14}B_{14}$ (labelled) is held
constant and initial spin period $P_0$ is varied,
while light dotted lines are for constant $P_0$ (labelled) and varying $B$.
Heavy dashed and dotted lines are the same but for a model which includes
magnetic dipole and GW energy loss, with r-mode amplitude $\alpha=0.1$.
Shaded region shows range of diffusion time which varies because
$\tdiff=\tdiff[\Erot(\Omegapeak)]$.
All models shown assume $\Esn=10^{51}\mbox{erg}$ and $M_{\rm ej}=5M_{\sun}$.
}
 \label{fig:peakr}
\end{figure}

It is clear that GW energy loss causes the neutron star to increase its
spin period more quickly at early times, and this decreases the amount
of rotational energy that can be used to power the SN/GRB light curve.
For lower magnetic fields, the effect of GWs is more dramatic, which can
be understood simply from the ratio of timescales given by
equations~(\ref{eq:tmagtgw}) and (\ref{eq:tmagtgwr}).
Some effects of GWs are illustrated in Figs~\ref{fig:peake} and
\ref{fig:peakr}, which plots peak photon luminosity $\Lpeak$
($\equiv$ maximum $\Lrad$) and time to peak luminosity $\tpeak\equiv t(\Lpeak)$.
The different lines of $\Lpeak(\tpeak)$ are calculated assuming either
constant magnetic field $B$ and varying initial spin period $P_0$ or
constant $P_0$ and varying $B$.
Similar plots are shown in \citet{kasenbildsten10} for the case of magnetic
dipole radiation only, and our results for this case are comparable.
Quantitative differences are due to a factor of two in the magnetic dipole
radiation timescale $\tmag$ (see equation~\ref{eq:tmag}), where
\citet{kasenbildsten10} assume $\sin^2\theta=1/2$ (see equation~\ref{eq:lmag})
and a factor of two and $\Erot(\Omega_0)$ in expansion velocity
and hence $\tdiff$ (see equation~\ref{eq:tdiff}).
We also denote the radiative diffusion timescale $\tdiff$
($\le 82\mbox{ d}$ for the parameters used here; see equation~\ref{eq:tdiff})
by the shaded region in Figs~\ref{fig:peake} and \ref{fig:peakr}.

For higher magnetic field strengths, magnetic dipole radiation is stronger,
and peak luminosities and times are the same when neglecting or including
the effect of GWs.
For the cases shown here ($\varepsilon=10^{-3}$ or $\alpha=0.1$),
$\tpeak$ shifts to later times and $\Lpeak$ is lower starting at
$B\sim 3\times 10^{14}\mbox{ G }(1\mbox{ ms}/P_0)$ for a GW ellipticity and
$B\sim 4\times 10^{14}\mbox{ G }(1\mbox{ ms}/P_0)^2$ for a GW r-mode
(see equations~\ref{eq:tmagtgw} and \ref{eq:tmagtgwr}, respectively).
These effects become significant at $B\lesssim 10^{14}\mbox{ G}$, and there
is no longer a one-to-one mapping between $P_0$--$B$ and $\tpeak$--$\Lpeak$.

\section{Discussion} \label{sec:discuss}

The theoretical model of converting rotational energy of a rapidly rotating
magnetar into magnetic dipole energy and this energy then powering ejecta of
SNe and GRBs has been successful in matching the observed light curve of SNe
and GRBs.
Here we consider the possible effects of GW emission on this magnetar model,
especially in light of the recent detection of GWs \citep{abbottetal16}.
We use the magnetar model to obtain simple constraints on the neutron
star ellipticity $\varepsilon$
(see also \citealt{laskyglampedakis16,moriyatauris16})
and r-mode oscillation amplitude $\alpha$,
since large values of $\varepsilon$ or $\alpha$ would cause a rapidly rotating
star to emit GWs and lose rotational energy at a faster rate than by magnetic
dipole radiation (see also \citealt{daietal16}).
We then account for GW emission processes within the magnetar model and show
how the evolution of the spin period and photon luminosity changes as a result
of inclusion of GW emission
(see also \citealt{kashiyamaetal16} for the case of $\varepsilon$).

Our constraint of $\varepsilon<0.01$ from Section~\ref{sec:magnetar}
and assumed value of $\varepsilon=10^{-3}$ in Section~\ref{sec:magnetargw}
are relatively large in magnitude.
GW detectors have thus far not detected GWs from rotating neutron stars,
with upper limits of $\varepsilon\sim 10^{-3}$ \citep{aasietal15,aasietal16}
and $\varepsilon\sim 10^{-6}-10^{-4}$ \citep{aasietal15,abbottetal16b} for
different frequency regimes (see also \citealt{aasietal14}).
Note that the neutron stars examined are much older than the magnetars
considered in the present work.
From a theoretical perspective, elastic deformations have maximum
$\varepsilon \sim 10^{-5}$ for neutron stars and $\sim 10^{-3}$ for more
exotic stars \citep{pitkin11,johnsonmcdanielowen13}.
Ellipticities created by strong magnetic fields can have
$\varepsilon\approx 10^{-6}$ (at a toroidal field strength of $10^{15}\mbox{ G}$
and increasing as $B^2$ or decreasing as $B$; \citealt{cutler02})
to $\varepsilon\sim 10^{-3}$ \citep{melatospriymak14,laskyglampedakis16}.
However the optimal geometry for strong GW emission, i.e. orthogonal magnetic
and rotation axes, may not occur even in the presence of extreme magnetic fields
(see \citealt{laskyglampedakis16}, for discussion; see also \citealt{lai01}).

Similarly to $\varepsilon$, the constraint of $\alpha<1$ from
Section~\ref{sec:magnetar} and assumed value of $\alpha=0.1$ in
Section~\ref{sec:magnetargw} are large amplitudes.
GW detectors have set upper limits in a wide range of $\alpha\sim 10^{-5}-0.1$,
depending on frequency \citep{aasietal15}.
X-ray observations yield upper limits of $\sim 10^{-6}$
\citep{mahmoodifarstrohmayer13}, as well as possible detection of r-modes
with amplitude $\alpha\sim 10^{-5}-10^{-3}$ in two neutron stars
\citep{strohmayermahmoodifar14,strohmayermahmoodifar14b,lee14}, although the
observed spin behaviour of one of these stars suggests surface phenomena which
would not generate GWs or impact the stellar spin rate \citep{anderssonetal14}.
But again these constraints are derived for much older neutron stars.
Note that we neglect evolution of the r-mode amplitude since $\alpha$ can reach
saturation at $10^{-3}$ or much lower \citep{arrasetal03,bondarescuwasserman13}
in $\lesssim 10^3\mbox{ s}$ \citep{owenetal98,alfordschwenzer14}
at $B\le 10^{15}\mbox{ G}$ \citep{holai00} and thus its evolution is not
relevant for SNe but may be relevant for GRBs \citep{yuetal10,chengyu14}.
However it is important to keep in mind that there is great uncertainty in
our understanding of the physics of r-modes
(see \citealt{hoetal11,haskelletal12}, for discussion).

Finally, even with a large ellipticity, GWs produced by a newborn rapidly
rotating magnetar would be difficult to detect, unless the source is
particularly nearby (see \citealt{kashiyamaetal16} for SNe and
\citealt{laskyglampedakis16} for GRBs).  The same is true for a large
amplitude r-mode oscillation, where the GW strain is
$h\sim 10^{-24}(10\mbox{ Mpc}/d)$ and $d$ is source distance.

\section*{Acknowledgements}
WCGH thanks Nils Andersson for discussion and Bob Nichol, Szymon Prajs,
and the anonymous referee for helpful comments.
WCGH acknowledges support from the Science and Technology Facilities
Council (STFC) in the United Kingdom.

\label{lastpage}
\end{document}